\documentclass{article}

\usepackage{arxiv}
\usepackage[utf8]{inputenc}
\usepackage[T1]{fontenc}
\usepackage{url}
\usepackage{booktabs}
\usepackage{amsfonts}
\usepackage{nicefrac}
\usepackage{microtype}
\usepackage{graphicx}
\graphicspath{{./images/}}

\usepackage{amsmath, amssymb}
\usepackage{xcolor}
\usepackage{listings}
\usepackage[hidelinks]{hyperref}
\usepackage[english]{babel}
\usepackage{caption}
\usepackage{subcaption}
\usepackage{tikz}
\usepackage{placeins}  
\usetikzlibrary{shapes, arrows, arrows.meta, positioning, shadows, calc, fit}

\colorlet{geomcpPrimary}{blue!65!black}
\colorlet{geomcpFill}{blue!4}
\tikzset{
  geomcpBox/.style={draw=geomcpPrimary!55, fill=geomcpFill, rounded corners=2pt,
    text width=11.5em, minimum height=3.2em, align=center, drop shadow},
  geomcpArrow/.style={-Latex, very thick, draw=geomcpPrimary!70},
  geomcpGroup/.style={draw=geomcpPrimary!30, rounded corners=2pt}
}

\definecolor{codebg}{HTML}{F8FAFC}
\definecolor{codeframe}{HTML}{D1D5DB}
\definecolor{codekw}{HTML}{1D4ED8}
\definecolor{codestr}{HTML}{065F46}
\definecolor{codecom}{HTML}{6B7280}
\definecolor{codenum}{HTML}{64748B}
\lstdefinestyle{geomcp}{
  backgroundcolor=\color{codebg},
  basicstyle=\ttfamily\scriptsize,
  frame=single,
  rulecolor=\color{codeframe},
  framesep=4pt,
  xleftmargin=1.25em,
  xrightmargin=0.5em,
  aboveskip=0.8\baselineskip,
  belowskip=0.6\baselineskip,
  numbers=left,
  numberstyle=\scriptsize\color{codenum},
  numbersep=10pt,
  keywordstyle=\bfseries\color{codekw},
  commentstyle=\itshape\color{codecom},
  stringstyle=\color{codestr},
  showstringspaces=false,
  columns=fullflexible,
  upquote=true,
  breaklines=true,
  breakatwhitespace=true,
  tabsize=2,
  captionpos=b
}
\definecolor{jsondelim}{HTML}{9CA3AF}
\definecolor{jsonpunct}{HTML}{6B7280}
\lstdefinelanguage{json}{
  basicstyle=\ttfamily\scriptsize,
  showstringspaces=false,
  breaklines=true,
  moredelim=**[is][\color{gray}\itshape]{@}{@}
}

\lstdefinelanguage{markdown}{
  basicstyle=\ttfamily,
  showstringspaces=false,
  commentstyle=\color{gray},
  keywordstyle=\color{blue}
}

\title{GeoMCP: A Trustworthy Framework for AI-Assisted Analytical Geotechnical Engineering}

\author{
  Yared W. Bekele \\
  Rock and Soil Mechanics Group \\
  SINTEF Community, Trondheim, Norway \\
  \texttt{yared.bekele@sintef.no}
}

\date{}

\begin{document}
\maketitle

\begin{abstract}
Analytical methods underpin geotechnical engineering practice, yet their implementation remains fragmented across error-prone spreadsheets and opaque proprietary software. While Large Language Models (LLMs) offer transformative potential for streamlining engineering workflows, their statistical nature fundamentally conflicts with the strict determinism required for safety-critical calculations. Their tendency to hallucinate formulas, misinterpret units, or alter methodologies between sessions creates a critical trust gap. This paper introduces GeoMCP, a framework built to bridge this gap via a key insight: engineering methods should be represented as structured data, not embedded code. GeoMCP captures analytical methods as "method cards", declarative JSON files defining formulas, units, applicability limits, and literature citations. A constrained symbolic engine executes these cards with verified dimensional consistency, while structured "Agent Skills" guide LLMs to apply engineering judgment and orchestrate the analysis. By exposing these verified capabilities through the Model Context Protocol (MCP), GeoMCP shifts the role of the AI from an unreliable calculator to an intelligent orchestrator. Validated against an official JRC Eurocode~7 worked example, the framework demonstrates computational parity with traditional approaches while ensuring complete mathematical transparency. Ultimately, GeoMCP provides a blueprint for transitioning the industry from isolated legacy software to an interoperable, AI-ready ecosystem where engineers can leverage modern AI without surrendering professional responsibility.
\end{abstract}


\section{Introduction}

Analytical methods, for example Terzaghi's bearing capacity theory and Rankine's earth pressure equations, form the foundation of geotechnical engineering practice. However, the tools used to apply these methods remain fragmented and often opaque. Spreadsheets are ubiquitous but error-prone, and studies have shown that a significant proportion contain errors that are difficult to detect in their cell-based structure~\cite{panko2008spreadsheet}. Across existing software and scripts, analytical formulations are often embedded in interfaces or implementation details, making method-level verification and citation harder than they should be.

The emergence of Large Language Models (LLMs) compounds this challenge. LLMs can understand technical problems and generate code, suggesting that AI assistants could streamline engineering workflows. However, LLMs are unreliable for safety-critical calculations: they could hallucinate formulas, misinterpret units, and conflate methods. For instance, when asked to compute Meyerhof's bearing capacity, an LLM may generate code that uses the Terzaghi $N_\gamma = 2(N_q+1)\tan\phi'$ expression instead of Meyerhof's own depth- and shape-corrected formulation, producing a plausible but unintended result with no indication of error. The core challenge is therefore not whether LLMs can help engineers, they clearly can, but how to harness their capabilities while maintaining the transparency, reliability, and auditability that engineering practice demands. We argue that the solution lies in providing LLMs with access to curated, verified calculation tools rather than asking them to generate calculations from memory.

This paper introduces GeoMCP, a framework for analytical geotechnical calculations designed for AI-assisted engineering. GeoMCP is founded on a key insight: \textit{geotechnical methods should be represented as structured data, not embedded code}. Each analytical method is captured as an explicit, human-readable ``method card'', a JSON file declaring variables, units, equations, assumptions, applicability limits, and literature sources. A constrained evaluation engine built on SymPy (symbolic mathematics) and Pint (physical units) executes method cards with verified dimensional consistency and full transparency. By exposing these capabilities through the Model Context Protocol (MCP)~\cite{anthropic_mcp}, GeoMCP enables AI assistants to perform reliable calculations while maintaining traceability to authoritative sources.

This data-centric approach transforms the role of AI in geotechnical engineering. Instead of generating potentially flawed calculations, an AI assistant equipped with GeoMCP becomes an interface to verified methods. An engineer can pose a natural language question about their problem, including supporting documents and data, and the assistant invokes GeoMCP tools to return not just an answer but a complete calculation trace showing inputs, intermediate steps, and the literature source for each formula.

This paper makes the following contributions:

\begin{enumerate}
\item A declarative JSON schema for representing analytical geotechnical methods as structured, citable data objects that capture formulas together with their engineering context - assumptions, applicability limits, and literature citations.

\item A safe, deterministic evaluation engine providing sandboxed symbolic evaluation (SymPy), automated dimensional analysis (Pint), and iterative solvers for circular dependencies common in geotechnical analysis.

\item Integration with the Model Context Protocol (MCP) and Agent Skills standards, making curated methods accessible to AI assistants and other client applications through a standardized interface.

\item Validation against the official JRC (Joint Research Centre) Eurocode~7 worked examples, focusing on the bearing capacity of shallow foundations.

\item A growing catalog of method cards covering common geotechnical analyses, with discussion of broader implications for AI-assisted engineering practice.
\end{enumerate}

The remainder of this paper is organized as follows. Section~\ref{sec:related} reviews related work. Section~\ref{sec:methodology} presents the design philosophy and system architecture. Section~\ref{sec:implementation} details the implementation of method cards, the evaluation engine, Agent Skills, and the MCP interface. Section~\ref{sec:application} validates GeoMCP against the official JRC Eurocode~7 worked examples. Section~\ref{sec:discussion} discusses advantages, limitations, and implications for AI-assisted engineering. Section~\ref{sec:conclusion} concludes with future directions.

\section{Related Work}
\label{sec:related}

GeoMCP draws upon and contributes to three interconnected research domains: computational tools for geotechnical engineering, artificial intelligence applications in civil engineering, and knowledge representation frameworks for engineering design. This section situates our work within these contexts and clarifies how GeoMCP addresses gaps in existing approaches.

\subsection{Geotechnical Software and Computational Tools}

Computational tools for analytical geotechnical calculations have evolved from simple chart look-ups and hand calculations to sophisticated commercial software suites that implement bearing capacity theories, settlement analyses, earth pressure calculations, and design code checks behind graphical user interfaces. These tools have become indispensable in practice, offering extensive material libraries, automated report generation, and thoroughly verified implementations that engineers rightly trust for routine design work. Yet two limitations persist. First, the specific analytical formulations employed are typically not easily traceable to their literature sources within the software itself: an engineer using a commercial tool to compute bearing capacity may not be able to confirm, from within the tool, exactly which published expression is being evaluated or which variant of a shape factor is applied. Second, these tools were designed for direct human operation through dedicated interfaces, not for programmatic or AI-assisted access. As LLMs and AI assistants become increasingly relevant to engineering workflows, the absence of machine-readable, standards-based interfaces to the underlying analytical methods represents a growing gap.

Alongside commercial software, a large share of day-to-day analytical geotechnical work is carried out with in-house spreadsheets, VBA macros, and bespoke scripts developed within individual firms or project teams. These internal tools are highly tailored and often efficient for the tasks they were built for, but they compound the limitations noted above. Formulas are embedded in cell references or procedural code with little or no connection to the published sources they implement; version control is rarely systematic; unit handling depends on the diligence of the original author; and the resulting calculations are difficult to audit, reproduce, or share beyond the team that created them~\cite{panko2008spreadsheet}. Like commercial packages, these tools are also not designed for integration into automated or AI-assisted workflows.

In the open-source domain, several Python libraries address different aspects of geotechnical analysis. \texttt{groundhog}~\cite{groundhog} is a broad-scope library covering site investigation, pile design, shallow foundation capacity, consolidation, earth pressure, and soil dynamics, with built-in parameter-range validation derived from the original source literature. \texttt{geolysis}~\cite{geolysis} focuses on bearing capacity, soil classification, and Standard Penetration Test analysis, while \texttt{pySlope}~\cite{pyslope} is purpose-built for slope stability assessment using Bishop's method of slices. These libraries make their implementations visible and modifiable, but they share a common architectural trait: formulas are encoded directly in program code, mixing calculation logic with implementation details. Verifying that an implementation matches the published method requires reading Python source; modifying or extending a method requires programming expertise; and citing the specific formula used in a calculation is not directly supported. Furthermore, these libraries are designed for programmatic use rather than integration with AI assistants or other client applications.

GeoMCP addresses these gaps by separating method definitions from evaluation logic. Each analytical method is captured as a self-contained JSON ``method card'' that declares its equations, variables, units, and, critically, the literature sources from which the formulas are drawn, making traceability a first-class property rather than an afterthought. Because the cards are plain data files, an engineer can inspect, verify, or extend a method using only domain knowledge and a text editor, without reading source code. The MCP interface then exposes these curated methods through a standardized protocol, enabling AI assistants and other client applications to discover and invoke verified calculations programmatically, a capability not offered by existing commercial, in-house, or open-source tools.

\subsection{Knowledge-Based Engineering and Formal Representations}

The concept of capturing engineering knowledge in machine-interpretable formats has a rich history in Knowledge-Based Engineering (KBE). Early expert systems, such as those developed for structural analysis and design in the 1980s, attempted to encode human expertise as rule-based systems. While technically impressive, these systems proved difficult to maintain, scale, and integrate with evolving engineering practice. The knowledge acquisition bottleneck, the challenge of eliciting and formalizing expertise from domain experts and encoding it in specialized knowledge representation languages, limited their adoption. Moreover, these systems often became brittle, failing to handle cases outside their training scope and requiring significant effort to update as codes and standards evolved.

More recent KBE research has focused on ontologies and semantic frameworks for representing engineering knowledge. Standards such as the Industry Foundation Classes (IFC), the Building Topology Ontology (BOT), and the Web Ontology Language (OWL) provide formal vocabularies for describing physical artifacts, their properties, and their relationships. These frameworks primarily address structural and topological descriptions rather than calculation procedures. They excel at representing ``what'' exists (a foundation has geometry, soil has properties, components relate to one another) but not ``how'' to compute derived quantities (bearing capacity from soil parameters using a specific analytical method). This gap motivates a complementary representation focused specifically on engineering calculations.

GeoMCP's method card concept shares structural characteristics with Minsky's frame-based knowledge representation~\cite{minsky1974framework}, in which structured objects encapsulate related data through typed slots. Each method card is a structured record containing slots for variables (with types, units, and roles), equations (expressed symbolically), applicability conditions, and literature sources. However, unlike classic AI frames, which embedded executable procedures in specialized languages, method cards use plain JSON and delegate evaluation to a constrained external engine. The resulting schema functions as a domain-specific language for analytical geotechnics: it defines a vocabulary and structure tailored to engineering calculations, enabling domain experts to contribute methods without programming expertise and addressing the knowledge acquisition bottleneck that limited earlier expert systems.

Because method cards are self-contained data files, they naturally support version control, peer review at the formula level, and community contribution through standard software development workflows. An engineer can fork a card, update an equation to match a revised design code, and submit the change for review, all without modifying source code. Each card's embedded literature citations ensure that the provenance of every formula remains explicit and machine-readable across revisions.

\subsection{Artificial Intelligence in Geotechnical Engineering}

The application of machine learning and artificial intelligence techniques to geotechnical problems has grown dramatically in recent years. Researchers have successfully applied artificial neural networks (ANNs) to predict soil behavior~\cite{shahin2001artificial}, support vector machines for settlement prediction~\cite{samui2008support}, and various ML algorithms for slope stability assessment~\cite{qi2010slope}, liquefaction potential evaluation, and bearing capacity estimation. These data-driven approaches can capture complex, nonlinear relationships in soil behavior and often provide good predictions when training data are representative. However, ML models fundamentally differ from physics-based analytical methods in their opacity and limited applicability outside training conditions. An ANN trained to predict bearing capacity does not provide insight into why a particular value results or how design changes would affect capacity. This black-box nature conflicts with engineering judgment and limits trust in safety-critical applications where understanding the reasoning behind a calculation is as important as the numerical result.

More recently, Large Language Models have demonstrated capabilities that could transform engineering practice: understanding technical specifications, generating code, and engaging in natural language dialogues. Several efforts have explored LLM-assisted engineering design, including code generation for preliminary design calculations. However, the reliability of LLM-generated engineering calculations remains a concern. Recent evaluation of LLMs on geotechnical problems revealed conceptual, calculation, and grounding errors across problem types, with unguided accuracy below 30\%~\cite{chen2024investigation}. Separately, work on tool-augmented language models~\cite{schick2024toolformer} suggests an alternative: rather than generating calculations, LLMs can learn to invoke external tools that perform them reliably. This insight motivates GeoMCP's architecture.

GeoMCP provides LLMs with verified tools rather than asking them to generate calculations. An LLM determines what calculation is needed, identifies appropriate method cards, supplies inputs, and interprets results, tasks suited to its natural language understanding, while delegating actual computation to deterministic, transparent method cards. This division leverages the strengths of both approaches: LLMs excel at understanding engineering context and translating queries into structured tool calls, while method cards ensure reliability and auditability.

\section{Methodology}
\label{sec:methodology}

This section describes the design philosophy and system architecture of GeoMCP. We begin by articulating the principles that guided our design decisions, then present the overall system architecture.

\subsection{Design Philosophy}

The central design principle of GeoMCP is that geotechnical methods should be represented as structured data, not embedded code. This principle addresses several fundamental requirements for engineering calculation tools in the era of AI assistance.

\textit{Transparency and auditability.} Engineering calculations, particularly those affecting safety, must be verifiable by competent professionals. When formulas are embedded within program code, whether in compiled commercial software or generated by LLMs, verification requires programming expertise and may be practically impossible when source code is proprietary or dynamically generated. By representing methods as explicit JSON documents, GeoMCP ensures that any engineer can inspect the formulas, compare them with published sources, and verify their correctness. This transparency is not merely a convenience but a professional responsibility: engineers must be able to explain and defend their calculations.

\textit{Citability and traceability.} Geotechnical methods have authoritative sources: textbooks, journal papers, design codes, and standards. A calculation's validity often depends on correctly implementing the method as published by a specific author or standard body. When methods are encoded in software, this connection to literature sources is typically lost or relegated to general documentation. Method cards make citation an integral part of the method representation: each card explicitly lists its literature sources, and individual formulas can reference specific equations from those sources. This enables engineers to trace every calculation back to its theoretical foundation, supporting both professional practice and educational use.

\textit{Extensibility and maintainability.} Geotechnical practice encompasses hundreds of analytical methods, each with multiple variants for different soil conditions, geotechnical structure types, and loading scenarios. A code-centric approach requires programming expertise to add new methods or modify existing ones, creating a bottleneck that limits who can contribute. The data-centric approach lowers this barrier: creating a method card requires geotechnical expertise and familiarity with JSON syntax, but not programming skills. This enables broader participation in expanding the method library and makes maintenance more sustainable as a community effort. When a design code is revised or a new variant is needed, updating a JSON file is more accessible than modifying source code.

\textit{AI readiness.} Large Language Models excel at understanding structured data and can reliably parse, query, and reason about JSON documents. By representing methods as structured data, we enable LLMs to understand what calculations are available, what inputs they require, and what assumptions apply, without requiring the LLM to generate calculation code. This division of responsibility, LLM for understanding context and selecting methods, deterministic engine for executing calculations, leverages the strengths of both approaches while maintaining reliability.

These principles led to a clear separation of concerns in the system architecture: method cards contain geotechnical knowledge (formulas, assumptions, citations), the evaluation engine provides safe computation, Agent Skills make it possible for LLMs to compose multiple cards into practical analysis procedures, and the MCP interface exposes these capabilities as tools. This separation ensures that improving one component (e.g., adding new methods, optimizing the engine, enhancing Agent Skills, or extending the interface) does not require changes to others.

\subsection{System Architecture}

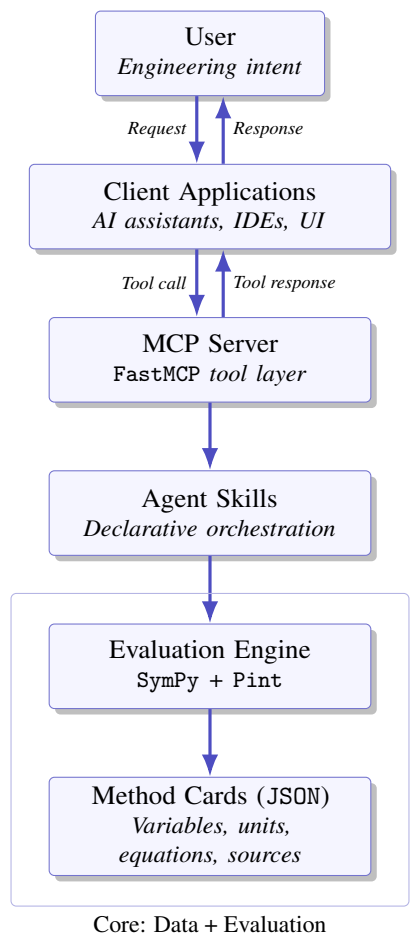
\begin{figure}[h!]
    \centering
    \begin{tikzpicture}[node distance=9mm]
        \node[geomcpBox, text width=8em] (user) {User\\{\footnotesize\textit{Engineering intent}}};
        \node[geomcpBox, below=of user, text width=13em] (client) {Client Applications\\{\footnotesize\textit{AI assistants, IDEs, UI}}};
        \node[geomcpBox, below=of client] (mcp) {MCP Server\\{\footnotesize\textit{\textup{\texttt{FastMCP}} tool layer}}};
        \node[geomcpBox, below=of mcp] (skills) {Agent Skills\\{\footnotesize\textit{Declarative orchestration}}};
        \node[geomcpBox, below=of skills] (engine) {Evaluation Engine\\{\footnotesize\textit{\textup{\texttt{SymPy}} + \textup{\texttt{Pint}}}}};
        \node[geomcpBox, below=of engine] (cards) {Method Cards (\texttt{JSON})\\{\footnotesize\textit{Variables, units, equations, sources}}};

        \draw[geomcpArrow] ([xshift=-5pt]user.south) -- node[left, font=\scriptsize\mdseries\itshape, text=black] {Request} ([xshift=-5pt]client.north);
        \draw[geomcpArrow] ([xshift=5pt]client.north) -- node[right, font=\scriptsize\mdseries\itshape, text=black] {Response} ([xshift=5pt]user.south);
        \draw[geomcpArrow] ([xshift=-5pt]client.south) -- node[left, font=\scriptsize\mdseries\itshape, text=black] {Tool call} ([xshift=-5pt]mcp.north);
        \draw[geomcpArrow] ([xshift=5pt]mcp.north) -- node[right, font=\scriptsize\mdseries\itshape, text=black] {Tool response} ([xshift=5pt]client.south);
        \draw[geomcpArrow] (mcp) -- (skills);
        \draw[geomcpArrow] (skills) -- (engine);
        \draw[geomcpArrow] (engine) -- (cards);

        \node[
            geomcpGroup,
            fit=(engine) (cards),
            inner xsep=5mm,
            inner ysep=4mm,
            label={[font=\small\mdseries, text=black]south:Core: Data + Evaluation}
        ] (core) {};
    \end{tikzpicture}
    \caption{GeoMCP system architecture showing a four-layer design: method cards as the knowledge base, a constrained evaluation engine for safe computation, Agent Skills for declarative orchestration and engineering reasoning, and an MCP server that exposes tools to clients through bidirectional request-response interactions.}
    \label{fig:architecture}
\end{figure}

GeoMCP follows a layered architecture designed to keep methods explicit, evaluation safe, and access flexible. The system comprises four distinct layers that interact with AI client applications, illustrated in Figure~\ref{fig:architecture}.

\textit{Method cards} form the knowledge base layer. Each card is a JSON file that completely specifies an analytical method: its variables (with physical units), equations (as SymPy expressions), variants (for different conditions), assumptions (applicability limits), and sources (literature citations). Cards are version-controlled, human-readable, and can be contributed by domain experts without programming expertise. All geotechnical knowledge, formulas, factors, conditions, resides in this layer. The deliberate choice of JSON rather than a specialized format ensures accessibility: any engineer can inspect, verify, or modify cards using standard text editors or version control tools.

\textit{The evaluation engine} provides the computation layer. Built on SymPy (symbolic mathematics) and Pint (physical units), it parses method card equations and evaluates them with strong safety and correctness guarantees. The engine uses a restricted SymPy expression allowlist, preventing code injection and ensuring reproducible evaluation. Pint provides automatic dimensional analysis, catching unit errors that are common sources of calculation mistakes. For methods requiring iterative solutions, the engine supports fixed-point iteration with convergence checking, with a growing list of capabilities. Importantly, the engine contains no geotechnical formulas itself, it is a general-purpose, safe evaluation framework for method cards.

\textit{Agent Skills} provide the orchestration and guidance layer. Rather than encoding multi-step analysis procedures as program code, GeoMCP captures them as structured Markdown documents following the Agent Skills specification~\cite{agentskills}. Each skill is organized as a directory containing a main instruction file (\texttt{SKILL.md}) with YAML frontmatter (name, description, version, category) and a \texttt{references/} subdirectory of companion documents. The instruction files follow a structured engineering reasoning approach tailored to the specific task, typically leading with problem understanding and site assessment \textit{before} any computation. For instance, a skill might instruct the assistant to first classify the problem type and assess site conditions, then select appropriate methods through engineering judgment, perform calculations through ordered tool calls, and finally interpret and sanity-check results against established professional rules of thumb. Reference files can provide domain knowledge that the AI assistant loads alongside the skill instructions, such as method comparison tables explaining when each method applies and why, typical parameter ranges for sanity checking, and relevant design code tables. Because skills are plain Markdown, domain experts can author, review, and extend them without programming expertise.

This approach addresses a fundamental limitation of tool-based AI interaction. Without guidance, AI models must infer multi-step analysis procedures and engineering reasoning from tool descriptions alone. In practice, this leads to suboptimal method selection, missing intermediate steps, or absent context. Implementing orchestration as hard-coded program logic is an alternative, but it does not scale: covering every geotechnical analysis domain would require thousands of lines of procedural code, each new domain requiring a software developer rather than a domain expert. Skills make orchestration explicit, auditable, and extensible by anyone with geotechnical expertise.

\textit{The MCP server} provides the interface layer. It exposes primitive tools for interacting with the method catalog (list methods, retrieve method cards, evaluate with numeric or unit-tagged inputs), skill discovery tools that enable clients to browse, search, and retrieve analysis skills, and composite tools for common multi-step analyses. The server's built-in instructions enforce a skill-first workflow: upon receiving a geotechnical question, the AI assistant first calls \texttt{recommend\_skills} to find the appropriate analysis skill, then loads it via \texttt{get\_skill} (with reference files), and follows the skill's engineering reasoning to evaluate the relevant method cards. This instruction-level enforcement ensures that AI assistants approach problems with structured engineering judgment rather than ad-hoc tool invocation. The server performs input validation and optional unit normalization, and returns deterministic results with intermediate values for traceability. By implementing the Model Context Protocol~\cite{anthropic_mcp}, GeoMCP becomes accessible to any MCP-compatible client, including AI assistants like ChatGPT or Claude, IDE plugins, and custom applications.

This architecture embodies our design principles. Methods remain explicit and auditable (cards layer), evaluation is safe and deterministic (engine layer), orchestration is declarative and domain-expert-authored (skills layer), and access is standardized (MCP server layer). An engineer can verify any calculation by inspecting the method card and re-computing with the published formulas, while an AI assistant follows a skill's engineering reasoning and invokes the same cards through MCP tools. Both paths execute identical underlying methods, ensuring that AI-assisted and manual analyses produce the same results. The following section describes the implementation of each layer.

\section{Implementation}
\label{sec:implementation}

This section describes the implementation of GeoMCP's core components: the method card schema that represents geotechnical knowledge, the evaluation engine that executes calculations safely and correctly, Agent Skills that provide declarative orchestration and engineering guidance, and the MCP server interface that exposes tools for method evaluation and skill discovery.

\subsection{Method Card Schema}

A method card is a structured JSON document that completely specifies an analytical geotechnical method. The schema is designed to capture not just the mathematics but the full engineering context needed to apply a method correctly and verify its implementation.

\textit{Core elements.} Each method card contains several essential sections. The \textit{metadata} section provides basic identification: a unique ID (e.g., \texttt{BEARING\_CAPACITY\_TERZAGHI}), human-readable title, category classification, and a descriptive overview. The \textit{variables} section declares all quantities used in the method, specifying for each its key (identifier used in equations), human-readable name, role (input, output, intermediate, or param for fixed constants), physical unit, and an optional detailed description. The \textit{variants} section allows a single card to encompass multiple versions of a method (e.g., for strip, square, or rectangular footings), each with its own set of equations. The \textit{equations} section defines the calculation procedure as a list of SymPy expressions, each targeting a specific variable and optionally including conditions for applicability and descriptive text. The \textit{assumptions} and \textit{applicability} sections formally outline theoretical assumptions and practical limits on when to use the method. Finally, the \textit{sources} section lists literature references and URLs, connecting the implementation to its theoretical foundation.

Table~\ref{tab:schema} summarizes the key schema fields. This structure ensures that method cards are self-documenting: an engineer can understand what a method calculates, what inputs it requires, under what conditions it applies, and where the formulas originate, all from the JSON file itself.

\begin{table}[h!]
\centering
\caption{Key fields in the method card schema. The schema defines a structured representation that captures both mathematical formulas and engineering context.}
\label{tab:schema}
\begin{tabular}{@{}lll@{}}
\toprule
Field & Type & Description \\
\midrule
\texttt{id} & string & Unique identifier (e.g., \texttt{BEARING\_CAPACITY\_MEYERHOF}) \\
\texttt{title} & string & Human-readable title \\
\texttt{category} & string & Classification (e.g., ``Shallow Foundations - Bearing Capacity'') \\
\texttt{description} & string & Brief overview of the method and its theoretical basis \\
\texttt{variables} & array & Variable specifications with roles, units, and descriptions \\
\texttt{variants} & array & Method variants for different calculation conditions \\
\texttt{assumptions} & array & Theoretical assumptions limiting the method's use \\
\texttt{applicability} & array & Practical guidelines on when to apply the method \\
\texttt{sources} & array & Literature citations and URLs connecting to references \\
\midrule
\multicolumn{3}{l}{\textit{Variable Specification}} \\
\midrule
\texttt{key} & string & Identifier used in equations (e.g., \texttt{phi\_prime}) \\
\texttt{name} & string & Human-readable variable name (e.g., \texttt{effective\_friction\_angle}) \\
\texttt{role} & string & \texttt{input}, \texttt{output}, \texttt{intermediate}, or \texttt{param} \\
\texttt{unit} & string & Physical unit (e.g., \texttt{kPa}, \texttt{radians}, \texttt{m}) \\
\texttt{description} & string & Plain language explanation of the specific quantity \\
\midrule
\multicolumn{3}{l}{\textit{Equation Specification}} \\
\midrule
\texttt{target} & string & Variable key this equation calculates \\
\texttt{sympy} & string & SymPy expression for calculation \\
\texttt{condition} & string & Optional applicability condition (e.g., \texttt{phi\_prime > 0}) \\
\texttt{description} & string & Optional text describing the specific equation or factor \\
\bottomrule
\end{tabular}
\end{table}

\textit{Design rationale.} The schema embodies several deliberate choices. JSON was selected for its ubiquity, any engineer can edit JSON files with standard text editors, and the format is supported by all major programming languages and version control systems. The schema is enforced at load time by Pydantic~\cite{pydantic}, a Python data-validation library that checks required fields, data types, and structural constraints, providing clear error messages before any card reaches the evaluation engine. Units are declared explicitly for every variable rather than assumed or documented separately, enabling automatic dimensional analysis via Pint~\cite{pint}. Multiple variants within a single card reduce duplication: common variables and metadata are defined once, while variant-specific equations adapt the method to different conditions. Literature sources are first-class elements rather than comments, making citations machine-readable and enabling automatic generation of reference lists in calculation reports.

Figure~\ref{fig:terzaghi_card} shows an abbreviated example of a method card for Terzaghi's bearing capacity theory. The card declares variables including the ultimate bearing capacity (\texttt{q\_ult}) as output and soil properties (\texttt{c\_prime}, \texttt{phi\_prime}, \texttt{gamma}) and geometry (\texttt{B}, \texttt{q}) as inputs. The variant for general shear failure in strip footings defines equations for the bearing capacity factors ($N_q$, $N_c$, $N_\gamma$) and the ultimate bearing capacity formula. The symbolic expressions use standard mathematical notation (e.g., \texttt{exp}, \texttt{tan}, \texttt{pi}) from the SymPy library, making them both machine-evaluable and human-readable.

\begin{figure}[h!]
\begin{lstlisting}[language=json, caption={Abbreviated method card for Terzaghi's bearing capacity theory showing the essential elements: metadata, variable declarations with units, symbolic equations, theoretical assumptions, and literature sources.}, label={fig:terzaghi_card}]
{
  "id": "BEARING_CAPACITY_TERZAGHI",
  "title": "Terzaghi's Bearing Capacity Theory",
  "category": "Shallow Foundations - Bearing Capacity",
  "description": "The first comprehensive theory for evaluating the ultimate bearing...",
  "variables": [
    {
      "key": "q_ult", "name": "ultimate_bearing_capacity", "role": "output",
      "unit": "kPa",
      "description": "Ultimate bearing capacity per unit area..."
    },
    {
      "key": "phi_prime", "name": "effective_friction_angle", "role": "input",
      "unit": "radians",
      "description": "Effective angle of internal friction..."
    },
    @"... (remaining variables omitted for brevity)"@
  ],
  "variants": [
    {
      "id": "general_shear_failure_strip",
      "title": "General Shear Failure for Strip Footings",
      "equations": [
        {
          "target": "N_q", "sympy": "exp(pi*tan(phi_prime))*tan(pi/4 + phi_prime/2)**2",
          "description": "Bearing capacity factor N_q"
        },
        {
          "target": "N_c", "sympy": "Piecewise(((N_q - 1)*cot(phi_prime), phi_prime > 0), (5.14, True))",          
          "description": "Bearing capacity factor N_c"
        },
        {
          "target": "N_gamma", "sympy": "2*(N_q + 1)*tan(phi_prime)",
          "description": "Bearing capacity factor N_gamma based on Terzaghi's..."
        },
        {
          "target": "q_ult", "sympy": "c_prime*N_c + q*N_q + 0.5*gamma*B*N_gamma",
          "description": "Ultimate bearing capacity for a continuous (strip) footing"
        }
      ]
    },
    @"... (remaining variants omitted for brevity)"@
  ],
  "assumptions": [
    "The foundation is shallow...",
    "The foundation base is rough...",
    @"... (remaining assumptions omitted)"@
  ],
  "applicability": [
    "Useful for preliminary estimates and for centrally loaded, shallow...",
    @"... (remaining applicability notes omitted)"@
  ],
  "sources": [
    {
      "title": "Terzaghi, K. (1943). Theoretical Soil Mechanics. John Wiley & Sons.",
      "url": @"... (URL omitted)"@
    },
    @"... (remaining sources omitted)"@
  ]
}
\end{lstlisting}
\end{figure}

This data-centric representation provides several practical advantages. First, the cards are \textit{verifiable}: an engineer can compare the SymPy expressions directly against published formulas to confirm correctness. Second, they are \textit{maintainable}: updating a method when a code is revised requires editing a single JSON file, not hunting through source code. Third, they are \textit{extensible}: adding new methods or variants requires no programming, enabling contributions from geotechnical specialists who are not software developers. Fourth, they are \textit{citable}: every calculation can trace back through the method card to literature sources, supporting professional documentation requirements.

\subsection{Evaluation Engine}

The evaluation engine executes calculations defined in method cards with guarantees of safety, dimensional correctness, and support for the iterative solution procedures common in geotechnical analysis.

\textit{Safe symbolic evaluation.} The engine is built on SymPy~\cite{meurer2017sympy}, a Python library for symbolic mathematics that represents expressions as algebraic objects rather than floating-point approximations. This means method card equations look like mathematical formulas (e.g., \texttt{exp(pi*tan(phi\_prime))}) and can be verified by inspection. The engine parses these expressions within a sandboxed environment that restricts available functions to a pre-approved allowlist of standard mathematical operations (\texttt{sin}, \texttt{cos}, \texttt{exp}, \texttt{sqrt}, \texttt{log}, etc.) and constants (\texttt{pi}, \texttt{e}). It does not use Python's \texttt{eval()} or allow access to built-in functions and modules, ensuring that method cards cannot execute arbitrary code, a critical security feature for a system that evaluates externally contributed logic.

\textit{Dimensional analysis.} Before evaluation, the engine uses Pint to verify that equations are dimensionally consistent: both sides of each equation must have compatible units, and operations must respect dimensional rules (e.g., pressure cannot be added to length). When an engineer supplies inputs with explicit units (e.g., ``32 degrees'' for friction angle), Pint automatically converts them to the units expected by the method card. If dimensional inconsistencies are detected, the engine raises a descriptive error before any calculation occurs, analogous to compile-time type checking in strongly typed programming languages.

\textit{Iterative solution support.} Many geotechnical methods involve circular dependencies where a variable cannot be isolated explicitly. For example, when sizing a foundation for bearing capacity, the allowable bearing pressure depends on the required footing width ($B$); however, $B$ itself is required to calculate the applied bearing pressure and satisfy the design criteria. The evaluation engine handles these cases through a multi-stage solving strategy. First, it attempts dependency resolution by making multiple passes through the equation set. For unresolved circular dependencies, the engine employs numerical fixed-point iteration with convergence checking. In general, the evaluation engine is continuously evolving in its capabilities to meet the expanding mathematical requirements of new method cards.

\textit{Performance and limitations.} The symbolic evaluation approach trades some computational performance for transparency and verifiability. For typical geotechnical design calculations involving dozens to hundreds of equations, evaluation time is on the order of milliseconds, fast enough for interactive use but slower than compiled numerical code. This is an acceptable trade-off for design calculations where engineer review time dominates computational time. For applications requiring thousands of repeated evaluations (e.g., Monte Carlo analysis), more optimized approaches might be needed, though the method card approach still provides value by ensuring the implementation matches published methods.

\subsection{Agent Skills}

While method cards capture individual analytical methods, practical engineering tasks typically require composing multiple methods/equations in sequence, selecting appropriate methods based on problem characteristics, and interpreting results against engineering judgment and professional norms. A traditional approach would encode these multi-step procedures as program code, but this creates a scalability barrier: each new geotechnical domain (earth pressure, settlement, slope stability, pile capacity) would require hundreds of lines of orchestration logic written by a software developer. GeoMCP addresses this through Agent Skills: structured Markdown documents that capture the same orchestration logic and engineering reasoning in a form that is both human-readable and machine-interpretable.

Each skill follows the Agent Skills specification~\cite{agentskills} and is organized as a directory containing a main instruction file (\texttt{SKILL.md}) with YAML frontmatter for metadata and Markdown prose for the analysis procedure, together with optional reference files that provide domain knowledge such as method comparison tables, typical value ranges, and examples. The AI assistant loads these reference files alongside the instructions when the skill is activated.

The key design decision in GeoMCP's skills is that they teach engineering \textit{reasoning}, not just tool invocation sequences. The shallow foundation bearing capacity skill, for example, provides structured guidance that mirrors how an experienced geotechnical engineer works. Rather than a rigid set of steps, the skill prompts the AI to first classify the problem type (design check vs. footing sizing) and assess site conditions (drainage, failure mode, groundwater). It then guides the selection of the appropriate method through a decision tree encoding engineering judgment, orchestrates calculations through ordered tool calls, executes design-code-specific workflows (e.g., EC7 partial factor design) when requested, and finally requires the assistant to interpret and sanity-check results against typical ranges and professional rules of thumb.

Critically, this reasoning phase occurs \textit{before any tool call}. The skill ensures the AI assistant stops to think about the problem, classifying it, assessing site conditions, and selecting the appropriate method, before reaching for a calculator. Without this guidance, AI models tend to skip directly to computation, often selecting methods without considering whether they are appropriate for the given conditions.

This approach provides several advantages over hard-coded orchestration. First, skills are \textit{extensible by domain experts}: a geotechnical engineer can author a new skill for, say, lateral earth pressure analysis using only Markdown, without writing Python code. Second, skills capture \textit{engineering reasoning} that code cannot: considerations such as ``Vesic is preferred over Terzaghi for rectangular footings with inclined loads because it accounts for compressibility and provides load inclination factors'' are naturally expressed in prose but awkward to encode as conditional logic. Third, skills are \textit{transparent and auditable}: the complete analysis procedure is visible in a human-readable document, enabling peer review by geotechnical specialists who may have no programming background.

\subsection{MCP Server Interface}

The MCP server provides a standardized interface through which client applications access GeoMCP capabilities. The Model Context Protocol (MCP), developed by Anthropic~\cite{anthropic_mcp}, defines how applications can expose their capabilities as discoverable, invokable tools with structured inputs and outputs. By implementing MCP via FastMCP (a Python library that handles protocol details), GeoMCP becomes compatible with any MCP-aware client, including Claude Desktop, ChatGPT, and custom applications, without requiring per-client integration code.

\textit{Tool categories.} The server exposes tools in several categories. \textit{Primitive tools} provide low-level access to the method catalog: listing methods, retrieving complete method cards for inspection, and evaluating methods with either normalized numeric inputs or unit-tagged inputs (e.g., ``30 deg'', ``18\,kN/m\textsuperscript{3}'') that are automatically converted to the card's expected units. Importantly, when an evaluation tool is invoked, the server returns an auditable trace of all inputs, intermediate variable calculations, and final results to the client. \textit{Skill discovery tools} enable AI assistants to browse, search, and retrieve analysis skills by relevance to a problem description. The server also provides session management for reusable parameter defaults and a diagnostic health check.

\textit{Skill-first workflow.} The server's built-in instructions guide AI assistants toward a skill-centric approach: when presented with a problem description, the assistant is prompted to use skill discovery tools to locate and load the appropriate analysis skill and its associated reference files. Rather than blindly invoking calculation tools, the assistant follows the domain-specific guidance provided by the skill. This instruction-level guidance ensures that the AI assistant approaches every problem with structured engineering reasoning rather than ad-hoc tool invocation.

\textit{Design decisions.} GeoMCP keeps the evaluation engine deterministic and side-effect free: given the same card, variant, and normalized inputs it produces the same result. For usability, the server supports an optional session context (defaults that can be reused across tool calls), but this convenience layer does not change the underlying mathematics, which always resides in method cards. This separation keeps the system auditable while supporting practical engineering analyses.

Because the protocol is standardized, improvements to MCP clients automatically benefit GeoMCP users, and additions to the method catalog or skill library become immediately available to all clients without client-side changes.

\section{Application and Validation}
\label{sec:application}

This section demonstrates GeoMCP from a user's perspective, showing how an AI assistant uses the framework to perform a complete Eurocode~7 bearing capacity design. We trace the full workflow, from natural language problem description through skill-guided analysis to validated results, against the official JRC (Joint Research Centre) worked example~\cite{jrc2005}. This walkthrough validates both the computational accuracy of the framework and the practical utility of skill-guided AI interaction.

\subsection{Method Catalog Overview}

The current GeoMCP implementation includes several method cards spanning basic soil mechanics, shallow and deep foundations, lateral earth pressure, settlement, and slope stability - with a growing number of methods and variants. For the purpose of validation in this paper, we focus on the bearing capacity subset, which includes established methods like Terzaghi's classical theory (\texttt{BEARING\_CAPACITY\_TERZAGHI})~\cite{terzaghi1943,das2018}, Meyerhof's shape- and depth-factored formulation (\texttt{BEARING\_CAPACITY\_MEYERHOF})~\cite{meyerhof1963,das2018}, Vesic's compressibility-aware approach (\texttt{BEARING\_CAPACITY\_VESIC})~\cite{vesic1973,das2018}, and Eurocode~7 drained and undrained calculations (\texttt{BEARING\_CAPACITY\_EUROCODE7})~\cite{en19971,simpson2012}.

\subsection{Validation Scenario: JRC Eurocode~7 Worked Example~A.3}

We validate GeoMCP against the official JRC Eurocode~7 worked examples~\cite{jrc2005}, published by the European Commission as the authoritative implementation guide for EN~1997-1.

The problem, as presented to the AI assistant, is a direct \textbf{design} request:

\begin{quote}
\textit{Design a strip foundation for a 4-story concrete building. The foundation has embedment depth $D_f = 1.5$\,m. Total characteristic permanent load from six columns is $G_{k,\mathrm{col}} = 3500.96$\,kN and characteristic variable load is $Q_k = 967.10$\,kN. The foundation length is $L = 21.4$\,m. Determine the required width $B$ for all Eurocode~7 Design Approaches.}
\end{quote}

\textbf{Data Consistency Note:} The JRC worked example text (p.105) states the groundwater level is at 1.5\,m.  However, the JRC geotechnical model Figure~A.3.111 (p.112) explicitly shows groundwater at 2.5\,m (1.0\,m below the base). Reproducing the specific JRC results requires using the Figure values.

\subsection{AI-Assisted Analysis Walkthrough}

\begin{figure}[h!]
    \centering
    \begin{tikzpicture}[
        actor/.style={font=\small\bfseries, text=geomcpPrimary},
        msg/.style={-Latex, thick, draw=geomcpPrimary!70},
        msgr/.style={Latex-, thick, draw=geomcpPrimary!70, dashed},
        note/.style={font=\scriptsize\itshape, text=black, align=left},
        phase/.style={font=\scriptsize\bfseries, text=geomcpPrimary, fill=geomcpFill, draw=geomcpPrimary!30, rounded corners=1pt, inner sep=2pt},
        timeline/.style={draw=geomcpPrimary!30, thick}
    ]
        \node[actor] (user) at (0, 0) {User};
        \node[actor] (ai) at (4.5, 0) {AI Assistant};
        \node[actor] (mcp) at (9, 0) {MCP Server};
        \node[actor] (cards) at (12.5, 0) {Method Cards};

        \draw[timeline] (0, -0.3) -- (0, -14.5);
        \draw[timeline] (4.5, -0.3) -- (4.5, -14.5);
        \draw[timeline] (9, -0.3) -- (9, -14.5);
        \draw[timeline] (12.5, -0.3) -- (12.5, -14.5);

        \node[phase] at (-1.8, -0.9) {Discovery};
        \draw[msg] (0, -1.2) -- node[above, font=\scriptsize] {Problem description} (4.5, -1.2);
        \draw[msg] (4.5, -1.8) -- node[above, font=\scriptsize] {\texttt{recommend\_skills}} (9, -1.8);
        \draw[msgr] (4.5, -2.3) -- node[above, font=\scriptsize] {Skill match} (9, -2.3);
        \draw[msg] (4.5, -2.8) -- node[above, font=\scriptsize] {\texttt{get\_skill} + refs} (9, -2.8);
        \draw[msgr] (4.5, -3.3) -- node[above, font=\scriptsize] {Instructions + references} (9, -3.3);

        \node[phase] at (-1.8, -3.9) {Reasoning};
        \node[note, text width=5cm] at (4.5, -4.5) {Classify problem, assess site,\\select EC7 Annex D method};

        \node[phase] at (-1.8, -5.5) {Computation};
        \draw[msg] (4.5, -5.9) -- node[above, font=\scriptsize] {\texttt{get\_ec7\_preset\_partials}} (9, -5.9);
        \draw[msgr] (4.5, -6.4) -- node[above, font=\scriptsize] {Partial factors} (9, -6.4);
        \draw[msg] (4.5, -7.0) -- node[above, font=\scriptsize] {\texttt{check\_footing\_uls\_ec7}} (9, -7.0);
        \draw[msg] (9, -7.5) -- node[above, font=\scriptsize] {Evaluate card} (12.5, -7.5);
        \draw[msgr] (9, -8.0) -- node[above, font=\scriptsize] {Results} (12.5, -8.0);
        \draw[msgr] (4.5, -8.5) -- node[above, font=\scriptsize] {$V_d$, $R_d$, utilization} (9, -8.5);
        \node[note] at (6.75, -9.2) {\textit{Adjust $B$, repeat until $V_d \approx R_d$}};

        \node[phase] at (-1.8, -10.0) {Interpretation};
        \node[note, text width=5cm] at (4.5, -10.7) {Compare DAs, verify factors,\\interpret economy};

        \draw[msgr] (0, -11.8) -- node[above, font=\scriptsize] {Design Report} (4.5, -11.8);

    \end{tikzpicture}
    \caption{Interaction sequence for skill-guided Eurocode~7 design. The assistant discovers and loads the appropriate skill, reasons about the problem, orchestrates an iterative search for the required width using the ULS check tool, and interprets the results before reporting.}
    \label{fig:sequence}
\end{figure}
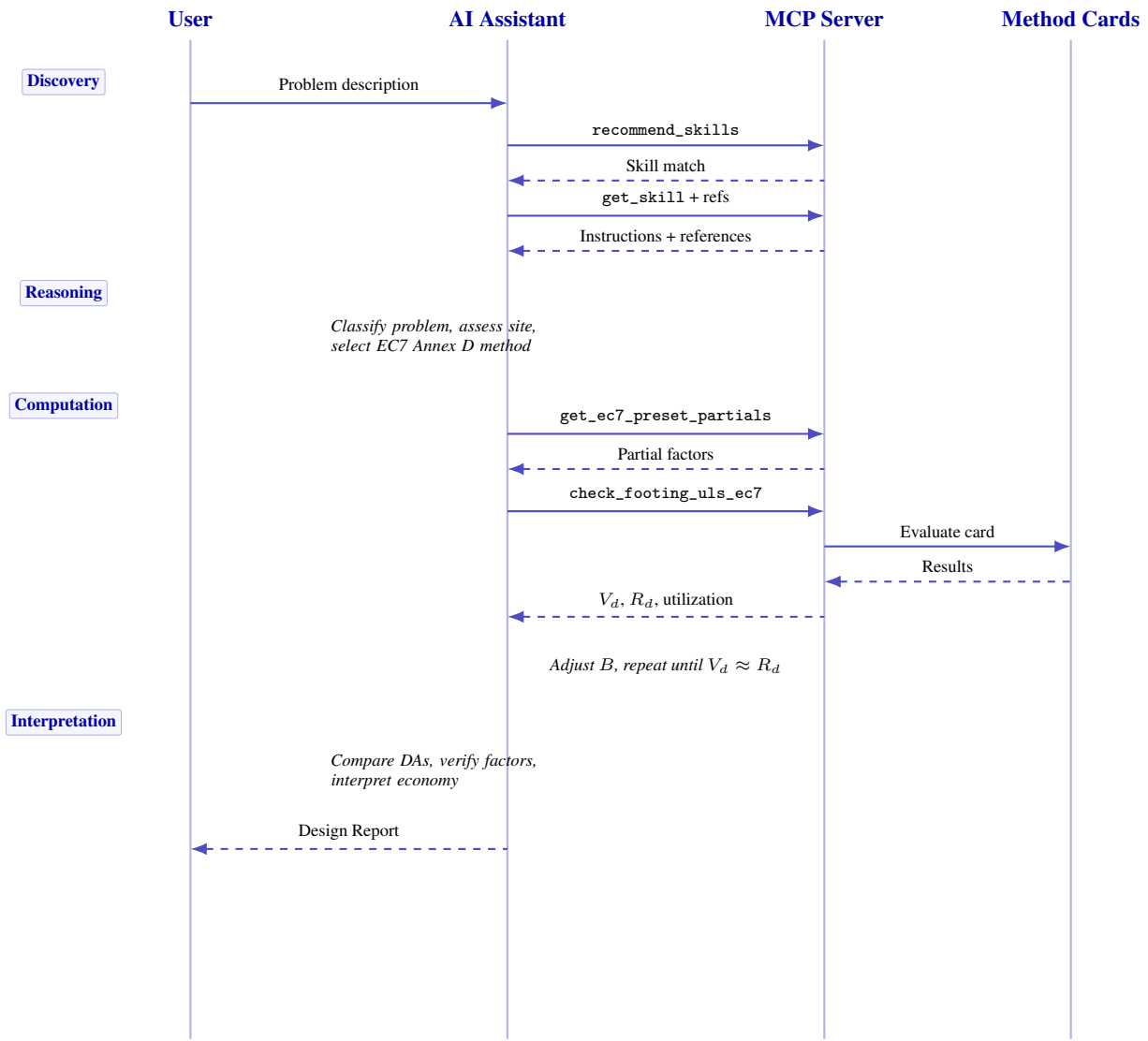

We trace the AI assistant's workflow as it solves for the required width $B$. Figure~\ref{fig:sequence} illustrates the interaction.

\textit{Discovery and reasoning.}
The assistant discovers the \texttt{shallow-foundation-bearing-capacity} skill and classifies the problem: ULS design for Eurocode~7. It notes that for Design Approach~1, two combinations must be checked (C1 and C2), while DA2 and DA3 have single combinations. Based on the soil properties ($\phi' = 38^\circ$), it anticipates general shear failure and selects the EC7 Annex~D method with drained conditions.

\textit{Design actions.}
Following the skill's suggested EC7 workflow, the assistant first retrieves the partial factors for each Design Approach. Listing~\ref{lst:partials} shows the tool call and response for DA1-C2. The partial factor set A2+M2+R1 ($\gamma_G = 1.0$, $\gamma_Q = 1.3$, $\gamma_\phi = 1.25$, $\gamma_R = 1.0$) is then applied to the characteristic loads to compute the design action $V_d$. Because the skill recommends checking all Design Approaches, the assistant repeats this for DA1-C1, DA2, and DA3.

\begin{lstlisting}[language=json, caption={MCP tool call and response for \texttt{geo\_get\_ec7\_preset\_partials}. The request contains only the Design Approach; the response returns the complete EN~1997-1 partial factor set.}, label=lst:partials]
// Request
{ "design_approach": "DA1-C2" }

// Response
{ "design_approach": "DA1-C2",
  "partials": {
    "gamma_G": 1.0,  "gamma_Q": 1.3,
    "gamma_phi": 1.25, "gamma_c": 1.25,
    "gamma_gamma": 1.0, "gamma_R": 1.0
  },
  "description": "Standard EN 1997-1 partial factors for DA1-C2" }
\end{lstlisting}

\textit{Bearing capacity and resistance.}
For each Design Approach, the assistant iteratively searches for the width $B$ where $V_d \approx R_d$. At each trial width, it calls \texttt{check\_footing\_uls\_ec7} with the geometry, characteristic soil parameters, Design Approach, and factored actions (including foundation self-weight at the trial width). The tool applies the partial factors, reduces the characteristic soil parameters to design values, evaluates the Eurocode~7 Annex~D bearing capacity method card, and returns an auditable trace with the utilization ratio.

\textit{Verification and interpretation.}
The assistant inspects the bearing capacity factors at the converged width and compares the Design Approach ordering (DA2 most economical, DA3 most conservative) against the JRC conclusions. It notes that settlement has not been checked, a serviceability limit state verification that would require additional method cards.

\subsection{Validation Results}

{\setlength{\textfloatsep}{8pt plus 2pt minus 2pt}
\setlength{\floatsep}{6pt plus 2pt minus 2pt}
\setlength{\intextsep}{6pt plus 2pt minus 2pt}

We validate GeoMCP's results against the JRC reference at three levels: bearing capacity factors, required foundation widths, and design actions/resistances at the JRC reference widths.

\paragraph{Bearing Capacity Factors}

Table~\ref{tab:jrc_factors} breaks down the intermediate bearing capacity and shape factors for DA1-C2 at the converged width. All factors agree with the worked example, which confirms that the \texttt{BEARING\_CAPACITY\_EUROCODE7} method card correctly implements the theory.

\begin{table}[!ht]
\caption{Bearing capacity and shape factors for DA1-C2 ($\phi'_d = 32.0^\circ$, $B = 1.497$\,m).}
\label{tab:jrc_factors}
\centering
\small
\begin{tabular}{@{}lrr@{}}
\toprule
\textbf{Parameter} & \textbf{JRC} & \textbf{GeoMCP} \\
\midrule
$N_q$ & 23.17 & 23.19 \\
$N_c$ & 35.48 & 35.51 \\
$N_\gamma$ & 27.72 & 27.74 \\
$s_q$ & 1.04 & 1.037 \\
$s_\gamma$ & 0.98 & 0.979 \\
\bottomrule
\end{tabular}
\end{table}

\paragraph{Required Foundation Widths}

The primary validation metric is the \textbf{required foundation width ($B$)}, as this single decision variable aggregates all partial factors, bearing capacity equations, shape factors, and load combinations. Table~\ref{tab:design_widths} compares the widths from the skill-orchestrated design search against the JRC reference.

\begin{table}[!ht]
\caption{Required foundation width for each Design Approach. DA1-C1 is not governing and is not reported in the JRC reference.}
\label{tab:design_widths}
\centering
\small
\begin{tabular}{@{}llrr@{}}
\toprule
& & \multicolumn{2}{c}{\textbf{Width $B$ (m)}} \\
\cmidrule(lr){3-4}
\textbf{Design Approach} & \textbf{Partial Sets} & \textbf{JRC} & \textbf{GeoMCP} \\
\midrule
DA1-C1 & A1+M1+R1 & -- & 1.038 \\
DA1-C2 & A2+M2+R1 & 1.50 & 1.497 \\
DA2 & A1+M1+R2 & 1.21 & 1.211 \\
DA3 & A1+M2+R3 & 1.74 & 1.738 \\
\bottomrule
\end{tabular}
\end{table}

The results confirm the JRC conclusion: DA2 is the most economical ($B = 1.211$\,m), while DA3 is the most conservative ($B = 1.738$\,m). The small differences in the third decimal place arise from minor rounding in the bearing capacity factors (Table~\ref{tab:jrc_factors}) and are well below any precision relevant to foundation design.

\paragraph{Design Actions and Resistances}

As a final verification, Table~\ref{tab:final_verification} compares $V_d$ and $R_d$ at the JRC reference widths. Because $V_d$ includes the foundation self-weight ($\gamma_{\text{concrete}} \times B \times D_f \times L$), it depends on the chosen width and is therefore presented after the widths are established. The $V_d$ values match exactly (to the precision reported), and $R_d$ agreement is within $0.012\%$, confirming that GeoMCP reproduces the JRC's complete computation chain.

\begin{table}[!ht]
\caption{Design actions ($V_d$) and resistances ($R_d$) at JRC reference widths.}
\label{tab:final_verification}
\centering
\small
\begin{tabular}{@{}lrrrr@{}}
\toprule
& \multicolumn{2}{c}{\textbf{$V_d$ (kN)}} & \multicolumn{2}{c}{\textbf{$R_d$ (kN)}} \\
\cmidrule(lr){2-3} \cmidrule(lr){4-5}
\textbf{DA} & \textbf{GeoMCP} & \textbf{JRC} & \textbf{GeoMCP} & \textbf{JRC} \\
\midrule
DA1-C2 & 5661.00 & 5661.00 & 5675.89 & 5676.52 \\
DA2 & 7160.11 & 7160.10 & 7151.43 & 7150.58 \\
DA3 & 7590.75 & 7590.75 & 7611.23 & 7612.07 \\
\bottomrule
\end{tabular}
\end{table}

All validations are implemented as automated regression tests in the repository, preventing drift as the method catalog evolves.

}

\section{Discussion}
\label{sec:discussion}

\subsection{Interpreting the Validation Results}

The JRC Eurocode~7 validation demonstrates that GeoMCP's evaluation engine reproduces an authoritative reference to high precision: bearing capacity factors within $0.15\%$, design actions exact to the reported precision, and design resistances within $0.012\%$. These results confirm that the fundamental architectural premise of GeoMCP is sound: abstracting geotechnical methods out of imperative code and representing them as symbolic expressions in declarative JSON introduces no meaningful numerical error. The framework achieves computational parity with established implementation approaches while simultaneously unlocking full method-level transparency and traceability.

\subsection{Implications for Engineering Practice}

The separation of method specification from evaluation logic has a practical consequence that extends beyond the AI use case: it makes the analytical methods themselves reviewable, versionable, and citable as independent artifacts. An engineer reviewing a GeoMCP calculation can trace every output back to a specific equation in a specific method card, and from there to the original published source. This level of traceability is exceedingly difficult to achieve with spreadsheets, where formulas are obscured behind cell references, or with proprietary software, where the underlying implementation is a compiled black box. 

This transparency is the prerequisite for responsibly deploying AI in geotechnical workflows. As the industry debates how professional liability applies when an AI agent selects a method or configures an analysis, the minimum requirement for a defensible workflow is auditability. By recording every decision in the analysis chain, GeoMCP ensures that transparency does not replace professional judgment, but rather makes it possible to exercise it effectively over an AI's actions.

Ultimately, this accountability unlocks a broader paradigm shift. We argue that the next wave of productivity gains in civil engineering will not come from individual Large Language Models becoming incrementally smarter, nor from isolated software tools adding more proprietary features. The future of engineering software lies in making our collective knowledge and trusted computational tools broadly interoperable and ``AI-ready.'' By extracting analytical methods from embedded code and exposing them as structured data through standardized protocols like MCP, GeoMCP transitions the industry away from building monolithic tools for human operators. It points toward an interconnected ecosystem where AI assistants can reliably orchestrate deterministic, verifiable engineering computations at scale.

\subsection{Limitations and Risks}

The JRC validation demonstrates precision within a specific problem class. A production-ready release of GeoMCP will require systematic validation across a broader range of geotechnical problems addressable via analytical methods. The availability of comprehensive verification datasets, complete with intermediate calculation values, is a necessity for this expansion. Furthermore, the method card approach is inherently restricted to closed-form equations and iterative systems.

A more critical limitation concerns the human-AI interaction paradigm. While GeoMCP could force an LLM to be deterministic with respect to use of analytical methods, it does not eliminate the potential for reasoning or data extraction errors. The LLM acts as the ``operator'' and is responsible for interpreting the problem scenario, selecting the appropriate skill, and extracting input parameters from the user's prompt. Errors in this interpretation phase, such as misidentifying soil parameters or incorrectly assessing drained versus undrained conditions, will propagate into the engine's inputs.

Although the system's auditable traces make all input values explicitly visible in the final report, this transparency introduces the secondary risk of automation bias. When a system presents a highly polished, professionally formatted design report containing academic citations and an irrefutable mathematical trace, the perceived authority of the output may discourage necessary critical review. Transparency provides the mechanism for verification, but organizational practices, such as mandatory independent checking, remain essential. The framework restricts the LLM's agency to reasoning and orchestration, strictly capping the potential for silent mathematical failure, but it reinforces the principle that AI tools amplify rather than replace professional engineering judgment.

\section{Conclusion}
\label{sec:conclusion}

The emergence of Large Language Models offers transformative potential for engineering workflows, but their statistical nature fundamentally conflicts with the strict determinism required for safety-critical structural and geotechnical design. This paper presented GeoMCP, a framework designed to bridge this gap, founded on the premise that engineering calculations should be represented as structured data rather than embedded code.

By capturing analytical methods as declarative JSON method cards, complete with formulas, units, applicability conditions, and literature citations, and evaluating them through a constrained symbolic engine, GeoMCP guarantees numerical parity with traditional implementation approaches. The validation against official Eurocode~7 worked examples confirmed that this data-centric abstraction introduces no numerical penalty, accurately reproducing authoritative design resistances and foundation dimensions while simultaneously providing a transparent, auditable calculation trace.

More importantly, the framework shifts the role of the AI from an unreliable calculator to an intelligent orchestrator. Through the Model Context Protocol and structured Agent Skills, an LLM dictates the engineering reasoning while delegating the math to a unified, verifiable source of truth. 

The architecture of GeoMCP demonstrates a viable path forward for the profession. It moves away from opaque spreadsheets and proprietary black boxes, advocating instead for an open, interoperable, and AI-ready ecosystem. By making computations explicit, citable, and accessible, GeoMCP allows engineers to leverage the vast capabilities of modern AI without surrendering their professional responsibility.

The source code for GeoMCP will be made publicly available on GitHub after the publication of this paper.

\subsubsection*{AI Use Disclosure}
The author declares that generative AI tools were used solely for proofreading, editing, and formatting assistance during the preparation of this manuscript. The research design, conceptualization, methodology, analysis, and interpretation of results represent the original work of the author.

\bibliographystyle{unsrt}
\bibliography{references}

\end{document}